# An Extreme Toughening Mechanism for Soft Materials

Shaoting Lin[1], Camilo Duque Londono[1], Dongchang Zheng[1,2], Xuanhe Zhao[1,3*]

[1]Department of Mechanical Engineering, Massachusetts Institute of Technology, Cambridge, MA 02139, USA.

[2]CAS Key Laboratory of Mechanical Behavior and Design of Materials, Department of Modern Mechanics, University of Science and Technology of China, Hefei 230026, People's Republic of China

[3]Department of Civil and Environmental Engineering, Massachusetts Institute of Technology, Cambridge, MA 02139, USA

[*]To whom correspondence should be addressed. Email: zhaox@mit.edu

# An Extreme Toughening Mechanism for Soft Materials

**Abstract**

Soft yet tough materials are ubiquitous in nature and everyday life. The ratio between fracture toughness and intrinsic fracture energy of a soft material defines its toughness enhancement. Soft materials' toughness enhancement has been long attributed to their bulk stress-stretch hysteresis induced by dissipation mechanisms such as Mullins effect and viscoelasticity. With a combination of experiments and theory, here we show that the bulk dissipation mechanisms significantly underestimate the toughness enhancement of soft tough materials. We propose a new mechanism and scaling law to account for extreme toughening of diverse soft materials. We show that the toughness enhancement of soft materials relies on both bulk hysteric dissipation, and near-crack dissipation due to mechanisms such as polymer-chain entanglement. Unlike the bulk hysteric dissipation, the near-crack dissipation does not necessarily induce large stress-stretch hysteresis of the bulk material. The extreme toughening mechanism can be universally applied to various soft tough materials, ranging from double-network hydrogels, interpenetrating-network hydrogels, entangled-network hydrogels and slide-ring hydrogels, to unfilled and filled rubbers.

**Introduction**

Soft yet tough materials – mainly constituted of polymer networks – are ubiquitous in nature and everyday life, ranging from animal and plant tissues [1, 2] , to synthetic and natural elastomers [3, 4], to recently developed tough hydrogels including double-network hydrogels [5, 6], interpenetrating-network hydrogels [7], polyampholyte hydrogels [8], and slide-ring hydrogels [9]. For instance, while the Young's moduli of natural muscles [10], triple-network elastomers [11], and interpenetrating-network hydrogels [7] are below a few megapascals, their fracture toughness can reach up to 10,000 Jm$^{-2}$ – approximating that of tough steels [12]. Such high fracture toughness of soft materials is crucial for their mechanical integrity and robustness in nature and in engineering applications.

Fracture toughness of soft materials has been long attributed to two physical processes [13-15]: 1) scission of a layer of polymer chains on the crack tip, and 2) hysteric mechanical dissipation in the bulk material around the crack tip due to mechanisms such as Mullins effect and viscoelasticity. The first process defines the intrinsic fracture energy $\Gamma_0$, and the second process gives the bulk hysteric dissipation's contribution to fracture toughness $\Gamma_D^{bulk}$. Consequently, the total fracture toughness of the soft material $\Gamma$ can be expressed as $\Gamma = \Gamma_0 + \Gamma_D^{bulk}$, which is often named the bulk dissipation model [14, 16-20]. The fracture toughness $\Gamma$ of a soft material can be measured as the critical energies required to propagate a crack by a unit area in a material under monotonic loading in a fracture test [Fig. 1 (a) and (c)]. The fatigue threshold $\Gamma_0$ of a soft material can be measured as the critical energies required to propagate a crack by a unit area in a material under infinite cycles of loading in a fatigue test [Fig. 1 (b) and (c)]. Despite their high fracture toughness up to 10,000 Jm$^{-2}$ [3-8], the measured intrinsic fracture energy of soft materials is usually on the order of 10 to 100 Jm$^{-2}$ [18, 21]. Soft materials' toughness enhancement – defined as $\Gamma/\Gamma_0$ – has been long attributed to the bulk dissipation mechanisms such as Mullins effect and viscoelasticity [13-15, 17, 19, 20].

With a combination of experiments and theory, this letter shows that the bulk dissipation mechanisms significantly underestimate the toughness enhancement of soft tough materials. We present a new model and scaling law to account for an extreme toughening mechanism in diverse soft tough materials, which relies on both bulk hysteric dissipation, and near-crack dissipation due to mechanisms such as polymer-chain entanglement and strain-induced crystallization. Using polyacrylamide (PAAm)-alginate hydrogels as an example, we show that the bulk dissipation model underestimates the toughness enhancement of PAAm-alginate hydrogels up to 6.6 times. In contrast, our new model can quantitively predict the toughness enhancement of PAAm-alginate hydrogels across a wide range of bulk hysteresis. We further show that the extreme toughening mechanism can be universally applied to various soft tough materials, ranging from interpenetrating-network hydrogels [7], double-network hydrogels [5, 6], slide-ring gels [9], and entangled hydrogels [22, 23], to unfilled and filled rubbers [24-27].

**Results and Discussion**

**Extreme Toughening Model.** Figure 2 schematically illustrates the physical picture of the extreme toughening model. Considering a notched soft material subject to a tensile load, crack propagation in the material first requires the scission of a single layer of polymer chains on the crack path. The required mechanical energy for chain scission divided by the area of crack surface at undeformed state gives the intrinsic fracture energy $\Gamma_0$, following Lake-Thomas model [Fig. 2 (a) and (d)] [18]. As the crack propagates, the material in a process zone around the crack path experiences a loading-unloading process, which dissipates mechanical energy due to bulk hysteresis, following the bulk dissipation model [Fig. 2 (b) and (d)] [14]. The dissipated energy divided by the area of the crack surface at undeformed state contributes to the fracture toughness by $\Gamma_D^{bulk}$. In addition, if the material contains polymer-chain entanglements, the crack propagation also requires pulling out of chains and delocalized damage of chains adjacent to the crack path, which give the near-crack dissipation [Fig. 2 (c) and (d)] [28]. The dissipated energy divided by the area of crack surface at undeformed state further contributes to the fracture toughness by $\Gamma_D^{tip}$. Therefore, the total fracture toughness of a soft material can be expressed as

$$\Gamma = \Gamma_0 + \Gamma_D^{bulk} + \Gamma_D^{tip} \tag{1}$$

The term $\Gamma_D^{bulk}$ in Eq. (1) can be estimated by

$$\Gamma_D^{bulk} = U_D L_D \tag{2}$$

where $U_D$ is the mechanical energy dissipated per volume of the process zone and $L_D$ is an effective size of the process zone. $U_D$ is a measurable quantity defined as $U_D = \oint_1^{\lambda_{max}} S d\lambda$, where $S$ and $\lambda$ are stress and stretch of the material under monotonic loading, $\lambda_{max}$ is the maximum stretch at which the material fails under the pure-shear deformation. The effective size of the process zone $L_D$ can be estimated by the stress distribution profile around the crack tip.

Without loss of generality, we take the soft material as a neo-Hookean solid. For a neo-Hookean solid under pure-shear fracture test [Fig. 1(a)], the leading order of the nominal stress at a point near the crack tip scales as $S \propto \sqrt{\Gamma \mu / x}$, where $\mu$ is the shear modulus of the materials and $x$ is the distance from the point to the crack tip [29]. Further, given the maximum nominal stress that the material can reach under the pure-shear deformation is $S_{max}$, we can estimate the size of the process zone as

$$L_D \propto \Gamma \mu / S_{max}^2 \propto \Gamma / U_{max} \tag{3}$$

where $U_{max} \propto S_{max}^2 / \mu$ is the maximum mechanical work done on the material under the pure-shear deformation. A combination of Eqs. (2) and (3) leads to

$$\Gamma_D^{bulk} \propto \Gamma h_m \tag{4}$$

where $h_m = U_D / U_{max}$ is the maximum stress-stretch hysteresis of the bulk material under the pure-shear deformation. A combination of Eqs. (1) and (4) further leads to

$$\frac{\Gamma}{\Gamma_0 + \Gamma_D^{tip}} = \frac{1}{1 - \alpha h_m} \tag{5}$$

where $0 \leq \alpha \leq 1$ is a dimensionless number depending on the stretch-dependent hysteresis of the bulk materials ($\alpha = 1$ for highly stretchable materials) [14, 30]. We further introduce a dimensionless parameter $\beta = (\Gamma_0 + \Gamma_D^{tip}) / \Gamma_0 \geq 1$ to account for the near-crack dissipation due to chain entanglements. Then, we can derive a governing equation for the toughness enhancement of soft tough materials as

$$\frac{\Gamma}{\Gamma_0} = \frac{\beta}{1-\alpha h_m} \tag{6}$$

When $h_m = 0$, Eq. (6) reduces to $\Gamma/\Gamma_0 = \beta$, corresponding to toughening of soft materials by the near-crack dissipation. When $\beta = 1$, Eq. (6) reduces to $\Gamma/\Gamma_0 = 1/(1-\alpha h_m)$, which recovers the bulk dissipation model.

**Materials.** We chose polyacrylamide-alginate (PAAm-alginate) hydrogels as a model material to validate the model. Due to its extremely high fracture toughness, PAAm-alginate hydrogel has been intensively exploited as a key component for devices and machines with examples such as tough hydrogel bonding [31], soft robots [26], hydrogel bandage [32], acoustic metamaterials [33], ultrasound imaging [34], and living sensors [35]. A PAAm-alginate hydrogel is made of two interpenetrating polymer networks: covalently crosslinked long-chain PAAm network, and ionically-crosslinked short-chain alginate network. The covalently crosslinked long-chain PAAm network provides the material's stretchable elasticity, and the ionically-crosslinked short-chain alginate network dissociates as the material is highly deformed, giving the material's bulk hysteresis. In this letter, we maintain the concentration and crosslinking density of PAAm network while varying the concentration and crosslinking density of alginate network, thereby tuning the bulk hysteresis of PAAm-alginate hydrogels.

We synthesize two series of PAAm-alginate hydrogels [Fig. 3(a)]. For both series of PAAm-alginate hydrogels, we start with preparing the pre-gel solution by dissolving the powders of sodium alginate (Sigma-Aldrich A2033) and acrylamide ($M_w = 71\,\text{g/mol}$, Sigma-Aldrich A8887) in deionized water. We fix the acrylamide concentration at 12 wt%, while varying the sodium alginate concentration $C_A$ from 0.0, 0.3, 0.6, 1.0, to 1.3 wt%. We add 110 μL 0.1 M ammonium persulfate (APS, Sigma-Aldrich A3678) as the thermal-initiator, 500 μL 0.23 wt% N,N'-Methylenebisacrylamide (MBAA, $M_w = 154\,\text{g/mol}$, Sigma-Aldrich 146072) as the crosslinker, and 20 μL N,N,N',N'-tetramethylethylenediamine (TEMED, Sigma-Aldrich T9281) as the crosslinking accelerator in 10 mL pre-gel solution, yielding solution A. The molar ratio between MBAA crosslinker and acrylamide monomer was fixed at 2263, which gives the average number of monomers between neighboring crosslinkers in the as-prepared state as $N = 2263$. Our rheology characterization has showed that a polymer network with $N = 2263$ synthesized from free radical polymerization leads to substantial chain entanglement in the polymer network [28]. Thereafter, we pour the solution A into a customized acrylic mold measuring $34 \times 6 \times 1.5$ mm$^3$. The mold is placed in a 50 °C oven to complete the thermal-induced free radical polymerization, giving the first series of PAAm-alginate hydrogels. The cured samples are further soaked in a bath of 0.01 wt% calcium chloride solution for 24 hours to induce ionic crosslinking between $Ca^{2+}$ and G unit of alginate chains, giving the second series of PAAm-alginate hydrogels. In the first series of PAAm-alginate hydrogels, the sodium alginate polymers are uncrosslinked mobile chains. In the second series of PAAm-alginate hydrogels, the sodium alginate polymers are ionically crosslinked into polymer networks. Unless otherwise stated, we denote the first series of PAAm-alginate hydrogels as hydrogels without $Ca^{2+}$ (i.e., W/O $Ca^{2+}$), and denote second series of PAAm-alginate hydrogels as hydrogels with $Ca^{2+}$ (i.e., W/ $Ca^{2+}$).

**Mechanical Characterizations.** We first characterize the stress-stretch curves of the two series of PAAm-alginate hydrogels up to failure points under the pure-shear deformation (see Supplemental Material). For hydrogels without $Ca^{2+}$, the sodium alginate concentration $C_A$ has little effect on the nonlinear stress-stretch relationship [Fig. S1(a)], because the alginate chains are uncrosslinked mobile chains and do not contribute to the elasticity of the hydrogels. In contrast, for hydrogels with $Ca^{2+}$, the sodium alginate concentration has significant impacts on stress-stretch curves [Fig. S1(b)]. As the sodium alginate concentration $C_A$ increases, the nominal stress increases accordingly while the ultimate stretch remains constant. Compared to hydrogels without $Ca^{2+}$, the ultimate stretches of hydrogels with $Ca^{2+}$ decrease drastically, possibly because the ionically-crosslinked alginate network suppresses the stretchablity of the polyacrylamide network.

We further characterize the stress-stretch hysteresis of the two series of PAAm-alginate hydrogels. Figure S2 plots the stress-stretch curves under one cycle of loading at different stretch levels for hydrogels without $Ca^{2+}$. The measured bulk hysteresis is consistently below 10% even when the maximum stretch approaches the failure points [Figs. 3(b) and 3(d)]. This is because the uncrosslinked alginate polymers do not contribute to elasticity or hysteresis of the material and the entangled PAAm polymer network exhibits low bulk hysteresis [28]. In contrast, since alginate polymers form the ionically-crosslinked network in hydrogels with $Ca^{2+}$, the alginate network unzips progressively when the material is highly deformed, which gives the huge bulk hysteresis

[Fig. 3(a)]. Figure 3(c) and Figure S3 plot the stress-stretch curves under one cycle of loading at different stretch levels for hydrogels with $Ca^{2+}$. The bulk hysteresis of hydrogels with $Ca^{2+}$ monotonically increases with the applied stretch and reaches a maximum plateau. We take the maximum plateau as the maximum bulk hysteresis $h_m$. As summarized in Fig. 3(d), the maximum bulk hysteresis of hydrogels with $Ca^{2+}$ increases with the alginate concentration $C_A$. This further indicates the critical role of ionically-crosslinked alginate network in promoting the bulk hysteresis.

We next use fracture and fatigue tests to measure the fracture toughness $\Gamma$ and fatigue threshold $\Gamma_0$ of the two series of PAAm-alginate hydrogels. We first adopt both pure-shear and single-notch methods to measure their fatigue thresholds (Figs. S4-S6), which gives their intrinsic fracture energies $\Gamma_0$. The measured fatigue thresholds of both series of hydrogels are consistently around 110 J/m². (Unless otherwise stated, the reported values of fatigue threshold have been converted to the corresponding values in the as-prepared or reference state by accounting for swelling of the hydrogels. The swelling ratios in volume are summarized in Fig. S7.) This indicates the presence of ionically-crosslinked alginate network does not contribute to the fatigue threshold [Fig. 3(b)], because the resistance to fatigue crack propagation after prolonged cycles of loading is the energy required to fracture a layer of PAAm polymer chains (i.e., the intrinsic fracture energy), which is unaffected by the additional bulk dissipation mechanisms by unzipping the ionically-crosslinked alginate network [36].

We further use the pure-shear method to measure the fracture toughness of the two series of PAAm-alginate hydrogels. For hydrogels without $Ca^{2+}$, the alginate concentration $C_A$ has little effect on the fracture toughness [Fig. S8]. Even though the bulk hysteric dissipations in hydrogels without $Ca^{2+}$ are negligible, the measured fracture toughness is still relatively high (480 J/m²), about 4.3 times of their fatigue threshold (i.e., 110 J/m²). This indicates that the difference between fracture toughness and fatigue threshold of hydrogels without $Ca^{2+}$ is due to the near-crack dissipation, not the bulk dissipation [28]. Therefore, the fracture toughness of hydrogels without $Ca^{2+}$ measures $\Gamma_0 + \Gamma_D^{tip}$ (Fig. 3). For hydrogels with $Ca^{2+}$, the alginate concentration $C_A$ significantly affects the fracture toughness (Fig. 4(a) and Fig. S8). As $C_A$ increases, the fracture toughness of hydrogels with $Ca^{2+}$ increases drastically from 500 to 2800 J/m² [Fig. 4(a)]. This enhancement of the fracture toughness is due to the bulk hysteric dissipation by unzipping the ionically-crosslinked alginate network; the level of bulk hysteric dissipation is determined by the alginate concentration $C_A$. Consequently, the fracture toughness of hydrogels without $Ca^{2+}$ measures $\Gamma_0 + \Gamma_D^{tip} + \Gamma_D^{bulk}$ (Fig. 4).

**Comparison between Experiments and Models.** Given the measured maximum bulk hysteresis $h_m$, fracture toughness $\Gamma$, and fatigue threshold $\Gamma_0$ of the hydrogels with $Ca^{2+}$, we summarize the measured toughness enhancement $\Gamma/\Gamma_0$ as a function of the measured maximum bulk hysteresis $h_m$ in Fig. 5(a). When the maximum bulk hysteresis is small, the toughness enhancement can still achieve 4.3. As the maximum bulk hysteresis increases, the toughness enhancement increases accordingly. When the maximum bulk hysteresis reaches 80%, the toughness enhancement can be as high as 22.

We then use Eq. (6) to calculate the relationship between fracture toughness enhancement $\Gamma/\Gamma_0$ and maximum bulk hysteresis $h_m$. The parameter $\beta = (\Gamma_0 + \Gamma_D^{tip})/\Gamma_0$ is identified as 4.3 given the measured $\Gamma_0 + \Gamma_D^{tip}$ (i.e., 480 J/m²) and the measured $\Gamma_0$ (i.e., 112 J/m²). The parameter $\alpha$ is taken as 1 since PAAm-alginate hydrogels are highly stretchable. Given the identified $\beta$ and $\alpha$, we can plot toughness enhancement $\Gamma/\Gamma_0$ as a function of the maximum bulk hysteresis $h_m$. As shown in Fig. 5(a), our extreme toughening model can quantitatively capture the toughness enhancement across a wide range of the maximum bulk hysteresis $h_m$. In contrast, we also plot $\Gamma/\Gamma_0$ versus $h_m$ following the bulk dissipation model, and the predicted toughness enhancement is significantly lower than the experimental results.

We further summarize reported toughness enhancement and maximum bulk hysteresis of various soft tough materials, including interpenetrating-network hydrogels [7, 21], double-network hydrogels [5, 37], entangled hydrogels [28], slide-ring gels [9], unfilled natural rubbers [24, 25], and filled styrene-butadiene rubbers [26, 27]. The predicted toughness enhancements following the bulk dissipation model are consistently lower than the measured values [Fig. 5(b)]. For example, the toughness enhancement of the interpenetrating-network hydrogels [7] with bulk hysteresis of around 80% should be around 5 following the bulk dissipation model, but the measured toughness enhancement is more than 20 [21]. The toughness enhancement of the double-network hydrogels [5] with bulk hysteresis of around 70% should be around 3.3 following the bulk dissipation model, but the measured toughness enhancement is at least 8 [37]. The toughness enhancement of unfilled natural rubbers with bulk hysteresis of around 20% should be around 1.2 following the bulk dissipation model [25], but the measured toughness enhancement is as high as 100 [24]. We envision our extreme toughening model can quantitatively capture the toughness enhancements of various soft toughen materials, because nearly all these soft tough materials contain substantial near-crack dissipation due to mechanisms such as chain entanglements.

**Conclusions**

In conclusion, we first use a combination of experiments and theory to show that the bulk dissipation mechanisms significantly underestimate the toughness enhancement of soft tough materials. We further propose a new mechanism and scaling law to account for an extreme toughening mechanism in diverse soft tough materials, which relies on both bulk hysteric dissipation, and near-crack dissipation due to mechanisms such as polymer-chain entanglement and strain-induced crystallization. Using polyacrylamide (PAAm)-alginate hydrogels as an example, we show that the bulk dissipation model underestimates the toughness enhancement of PAAm-alginate hydrogels up to 6.6 times. In contrast, our new model can quantitively predict the toughness enhancement of PAAm-alginate hydrogels across a wide range of bulk hysteresis. We envision the extreme toughening mechanism can be universally applied to various soft tough materials, ranging from double-network hydrogels, interpenetrating-network hydrogels, entangled-network hydrogels and slide-ring hydrogels, to unfilled and filled rubbers. Our study resolves a fundamental dilemma in toughening mechanisms of soft materials. It is hoped that this work can help lay the theoretical foundation for the development of next-generation tough, fatigue-resistant, and resilient soft materials.

**References**

[1] M. A. Meyers, J. McKittrick, P.-Y. Chen, science **2013**, 339, 773.
[2] W. Huang, D. Restrepo, J. Y. Jung, F. Y. Su, Z. Liu, R. O. Ritchie, J. McKittrick, P. Zavattieri, D. Kisailus, Advanced Materials **2019**, 31, 1901561.
[3] S. Toki, B. S. Hsiao, Macromolecules **2003**, 36, 5915.
[4] A. Gent, S. Kawahara, J. Zhao, Rubber Chemistry and Technology **1998**, 71, 668.
[5] J. P. Gong, Y. Katsuyama, T. Kurokawa, Y. Osada, Advanced materials **2003**, 15, 1155.
[6] Q. Chen, L. Zhu, C. Zhao, Q. Wang, J. Zheng, Advanced materials **2013**, 25, 4171.
[7] J.-Y. Sun, X. Zhao, W. R. Illeperuma, O. Chaudhuri, K. H. Oh, D. J. Mooney, J. J. Vlassak, Z. Suo, Nature **2012**, 489, 133.
[8] T. L. Sun, T. Kurokawa, S. Kuroda, A. B. Ihsan, T. Akasaki, K. Sato, M. A. Haque, T. Nakajima, J. P. Gong, Nature materials **2013**, 12, 932.
[9] C. Liu, N. Morimoto, L. Jiang, S. Kawahara, T. Noritomi, H. Yokoyama, K. Mayumi, K. Ito, Science **2021**, 372, 1078.
[10] D. Amiel, C. Frank, F. Harwood, J. Fronek, W. Akeson, Journal of Orthopaedic Research **1983**, 1, 257.
[11] E. Ducrot, Y. Chen, M. Bulters, R. P. Sijbesma, C. Creton, Science **2014**, 344, 186.
[12] R. O. Ritchie, J. F. Knott, J. Rice, Journal of the Mechanics and Physics of Solids **1973**, 21, 395.
[13] X. Zhao, Soft matter **2014**, 10, 672.
[14] T. Zhang, S. Lin, H. Yuk, X. Zhao, Extreme Mechanics Letters **2015**, 4, 1.
[15] R. Long, C.-Y. Hui, Soft Matter **2016**, 12, 8069.
[16] A. G. Evans, Journal of the American Ceramic society **1990**, 73, 187.
[17] R. McMeeking, A. Evans, Journal of the American Ceramic Society **1982**, 65, 242.
[18] G. Lake, A. Thomas, Proceedings of the Royal Society of London. Series A. Mathematical and Physical Sciences **1967**, 300, 108.
[19] J. Yang, R. Bai, B. Chen, Z. Suo, Advanced Functional Materials **2020**, 30, 1901693.
[20] P.-G. de Gennes, Langmuir **1996**, 12, 4497.
[21] R. Bai, Q. Yang, J. Tang, X. P. Morelle, J. Vlassak, Z. Suo, Extreme Mechanics Letters **2017**, 15, 91.
[22] D. Zheng, S. Lin, J. Ni, X. Zhao, Extreme Mechanics Letters **2022**, 101608.
[23] C. Norioka, Y. Inamoto, C. Hajime, A. Kawamura, T. Miyata, NPG Asia Materials **2021**, 13, 1.
[24] G. Lake, P. Lindley, Journal of Applied Polymer Science **1965**, 9, 1233.

**Acknowledgements**

This work was supported by National Institutes of Health (No.1R01HL153857-01) and National Science Foundation (No. EFMA-1935291).

**Author contributions**

S.L. and X. Z. conceived the idea, designed the study and interpreted the results. S.L., C. D. L., and D. Z. performed the experiments, analyzed the data, and interpreted the results. S.L. and X. Z. drafted the manuscript with inputs from all other authors. X.Z. supervised the study.

# Figures and Figure Captions

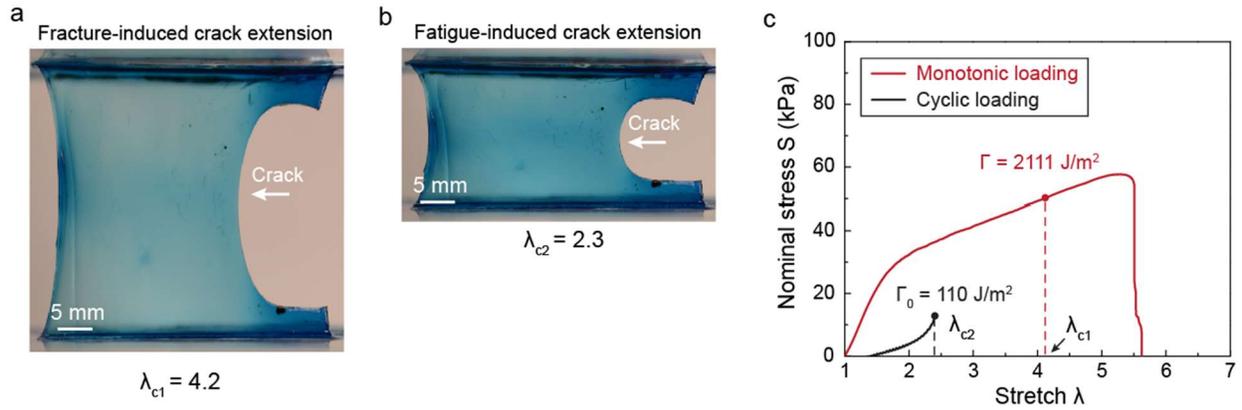

**Fig. 1. Fracture and fatigue tests of tough hydrogels measuring fracture toughness $\Gamma$ and fatigue threshold $\Gamma_0$.** (a) Image of fracture-induced crack extension in a tough hydrogel under monotonic loading, measuring the critical stretch $\lambda_{c1}$ for crack propagation. (b) Image of fatigue-induced crack extension in the same tough hydrogel under cyclic loading, measuring the critical stretch $\lambda_{c2}$ for crack propagation under almost infinite cycles of loading. (c) Nominal stress $S$ versus stretch $\lambda$ curves of the un-notched tough-hydrogel samples under monotonic loading and cyclic loading. The un-notched samples have the same material and dimensions as the samples in (a) and (b). Given the identified $\lambda_{c1}$, one can calculate the fracture toughness as $\Gamma = H\int_{1}^{\lambda_{c1}} S d\lambda = 2111 \text{ J/m}^2$. Given the identified $\lambda_{c2}$, one can calculate the fatigue threshold as $\Gamma_0 = H\int_{1}^{\lambda_{c2}} S d\lambda = 110 \text{ J/m}^2$. $H$ is the initial height of the sample.

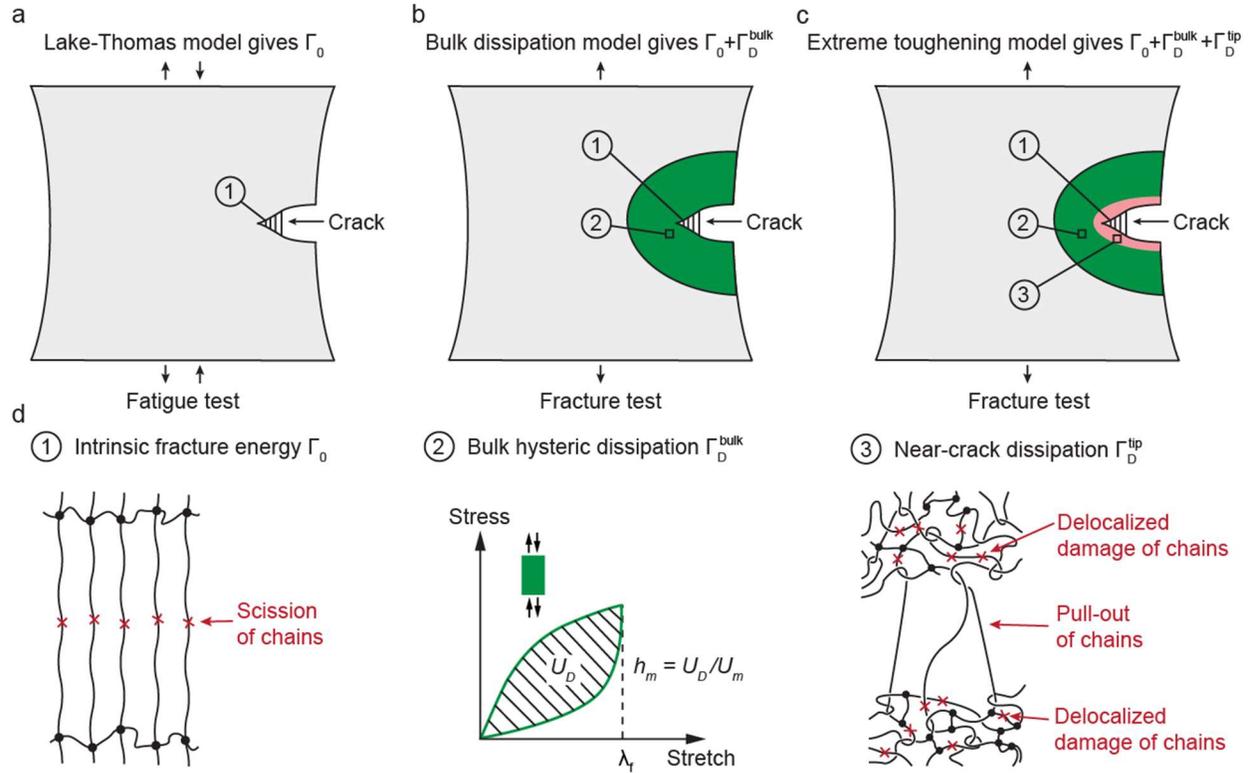

**Fig. 2. Schematic illustration of molecular mechanisms of three fracture models.** (a) Schematic illustration of the Lake-Thomas model accounting for intrinsic fracture energy of soft materials $\Gamma_0$. Intrinsic fracture energy of soft materials is typically measured by the fatigue test. (b) Schematic illustration of the bulk dissipation model accounting for two contributions to the total fracture toughness measured in the fracture test: intrinsic fracture energy $\Gamma_0$ and bulk hysteric dissipation $\Gamma_D^{bulk}$. (c) Schematic illustration of the extreme toughening model accounting for three contributions to the total fracture toughness measured in the fracture test: intrinsic fracture energy $\Gamma_0$, bulk hysteric dissipation $\Gamma_D^{bulk}$, and near-crack dissipation $\Gamma_D^{tip}$. (d) Schematic illustration of scission of a layer of chains for the intrinsic fracture energy $\Gamma_0$, large stress-stretch hysteresis loop for bulk hysteric dissipation $\Gamma_D^{bulk}$, and pull-out of chains and/or delocalized damage of chains for near-crack dissipation $\Gamma_D^{tip}$.

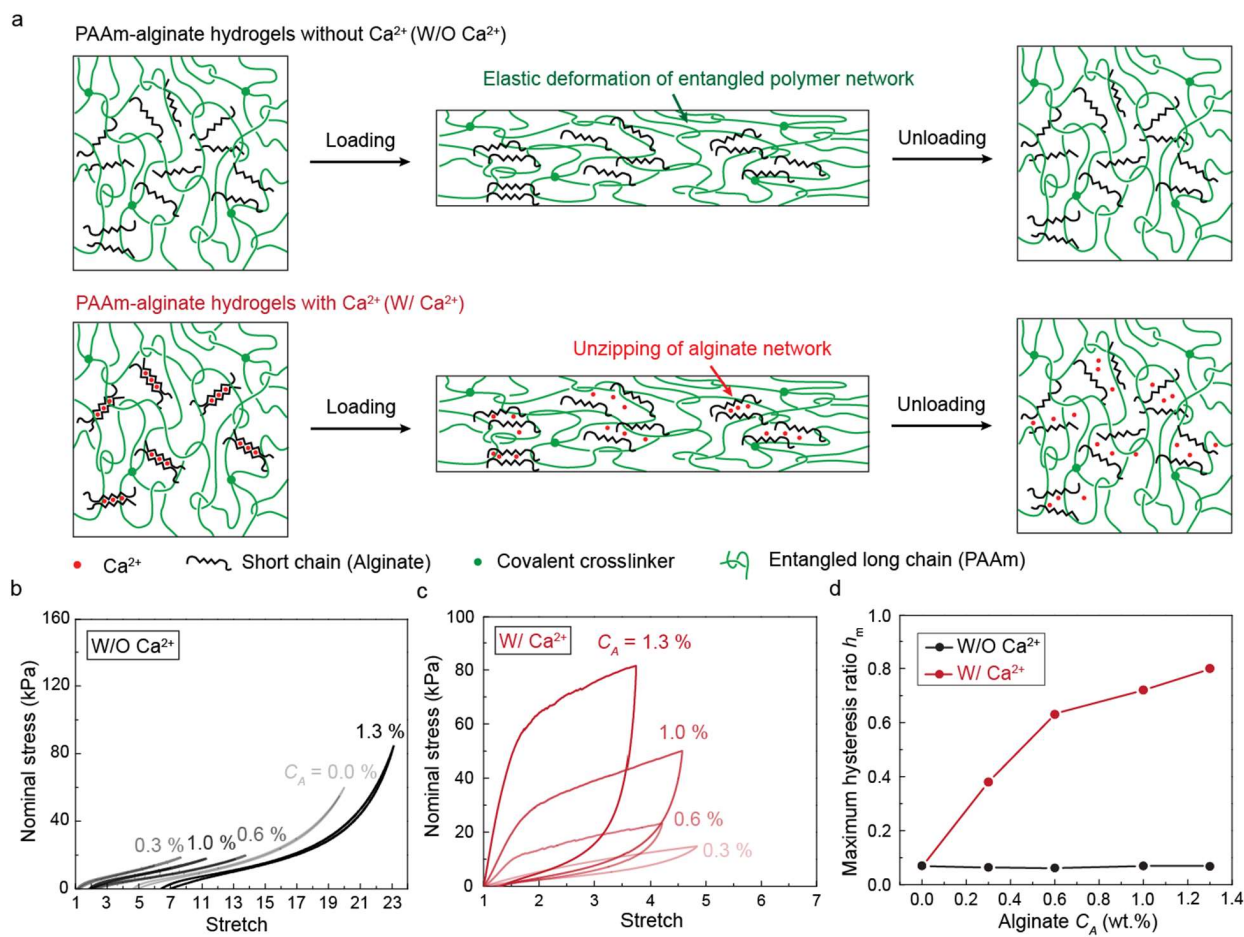

**Fig. 3. Stress-stretch hysteresis in hydrogels with and without Ca$^{2+}$.** (a) Schematic illustration of molecular pictures of the two series of PAAm-alginate hydrogels under a single cycle of loading and unloading. (b) Nominal stress versus stretch curves of hydrogels without Ca$^{2+}$ containing various alginate concentration $C_A$ under a single cycle of loading and unloading. (c) Nominal stress versus stretch curves of hydrogels with Ca$^{2+}$ containing various alginate concentration $C_A$ under a single cycle of loading and unloading. (d) The maximum hysteresis $h_m$ as a function of alginate concentration $C_A$ for the two series of PAAm-alginate hydrogels.

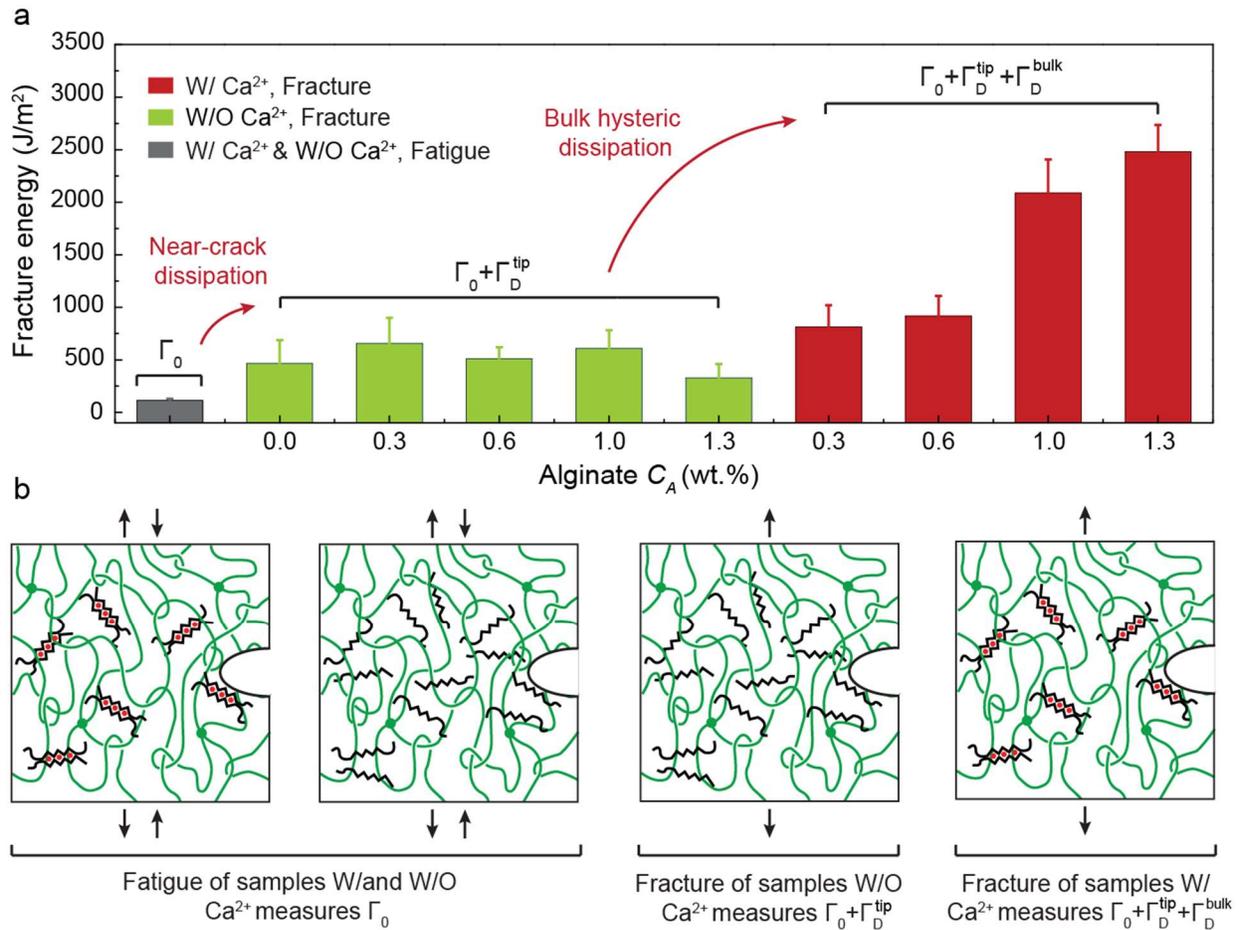

**Fig. 4. Summarized fracture toughness and fatigue threshold of hydrogels with and without $Ca^{2+}$.** (a) Three levels of fracture energies of the two series of PAAm-alginate hydrogels. (b) Schematic illustration of fatigue test of hydrogels with and without $Ca^{2+}$ measuring $\Gamma_0$, fracture test of hydrogels without $Ca^{2+}$ measuring $\Gamma_0 + \Gamma_D^{tip}$, and fracture test of hydrogels with $Ca^{2+}$ measuring $\Gamma_0 + \Gamma_D^{tip} + \Gamma_D^{bulk}$.

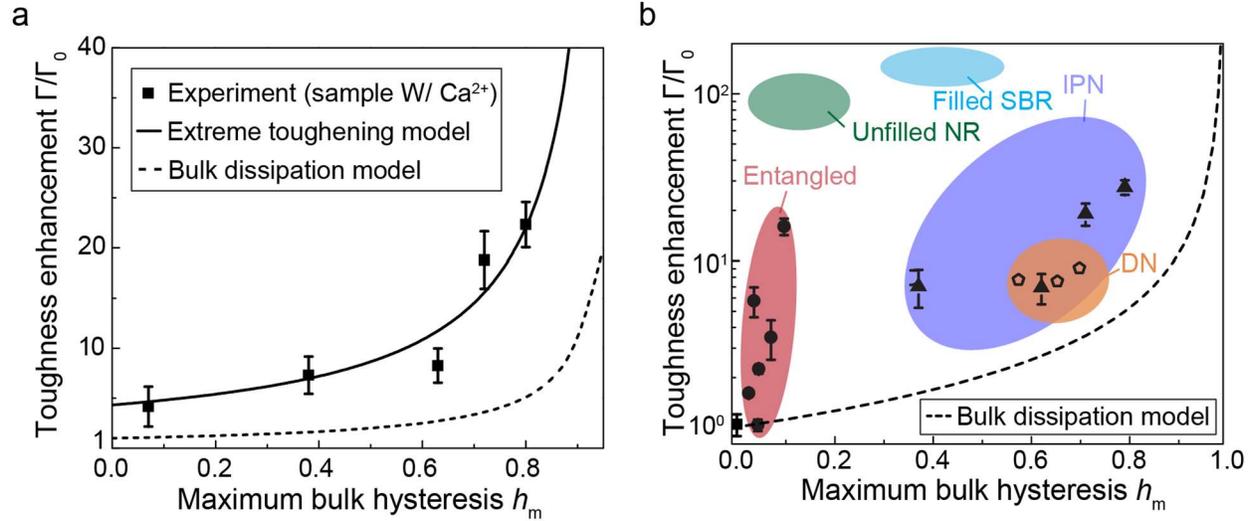

**Fig. 5. Comparisons between experiments and models for toughness enhancement versus maximum bulk hysteresis $h_m$.** (a) Comparisons of toughness enhancement $\Gamma/\Gamma_0$ versus maximum bulk hysteresis $h_m$ between the experimental results and the two models (extreme toughening model and bulk dissipation model). (b) Toughness enhancement $\Gamma/\Gamma_0$ and maximum bulk hysteresis $h_m$ for including interpenetrating-network (IPN) hydrogels [7, 21], double-network (DN) hydrogels [5, 37], entangled hydrogels [28], unfilled natural rubbers (NR) [24, 25], and filled styrene-butadiene rubbers (SBR) [26, 27]. The bulk dissipation model consistently underestimates the toughness enhancement of these soft tough materials.

Supplementary Information for

**An Extreme Toughening Mechanism for Soft Materials**

Shaoting Lin[1], Camilo Duque Londono[1], Dongchang Zheng[1], Xuanhe Zhao[1,2*]

[1]Department of Mechanical Engineering, Massachusetts Institute of Technology, Cambridge, MA 02139, USA

[2]Department of Civil and Environmental Engineering, Massachusetts Institute of Technology, Cambridge, MA 02139, USA

*Corresponding author: zhaox@mit.edu

**Materials.** Unless otherwise specified, the chemicals used in this work were purchased from Sigma-Aldrich and used without further modification. To synthesize the two series of PAAm-alginate hydrogels, we dissolved the powders of sodium alginate (Sigma-Aldrich A2033) and acrylamide ($M_w = 71 \text{ g/mol}$, Sigma-Aldrich A8887) in deionized water, yielding pre-gel solution with fixed acrylamide concentration at 12 wt% but varied alginate concentration from 0.0, 0.3, 0.6, 1.0, to 1.3 wt%. For each batch of sample, we further added 110 μL 0.1 M ammonium persulfate (APS, Sigma-Aldrich A3678) as the thermal-initiator, 500 μl 0.23 wt% N,N'-Methylenebisacrylamide (MBAA, $M_w = 154 \text{ g/mol}$, Sigma-Aldrich 146072) as the crosslinker, and 20 μL N,N,N',N'-tetramethylethylenediamine (TEMED, Sigma-Aldrich T9281) as the crosslinking accelerator in 10 mL of the pre-gel solution. Thereafter, we poured the solution into a customized acrylic mold measuring $34 \times 6 \times 1.5 \text{ mm}^3$. The mold was placed in a 50 °C oven to complete the thermal-induced free radical polymerization, resulting in the first series of PAAm-alginate hydrogels. The cured samples were further soaked in a bath of 0.01 wt% calcium chloride (Sigma-Aldrich C4901) solution for 24 hours to form ionic crosslinking between $Ca^{2+}$ and alginate chains, resulting in the second series of PAAm-alginate hydrogels.

**Chemically anchoring PAAm on glass fixtures.** To accurately capture nonlinear stress-stretch response of hydrogels up to material failures, we chemically anchored PAAm on glass fixtures. The glass slides were treated by oxygen plasma (30 W at a pressure of 400 mtorr, Harrick Plasma PDC-001) for 2 min. During oxygen plasma treatment, silicon oxide layers on glass slides react to hydrophilic hydroxyl groups by oxygen radicals produced by the oxygen plasma. After the plasma treatment, we immersed the glass slides in a bath of silane solution for 1 hour, which was prepared by mixing 500 mL deionized water, 50 μL of acetic acid with pH 3.5, and 2 mL silane 3-(trimethoxysilyl) propyl methacrylate (TMSPMA, Sigma-Aldrich 440159). The hydroxyl groups produced by oxygen plasma form hydrogen bonds with silanes in the silane solution. Thereafter, glass slides were washed with deionized water and dried using nitrogen gas. The silane treated glass slides were placed in the customized mold for sealing the pre-gel solution. During the curing of the pre-gel solution, a copolymerization also occurs between the methacrylate group in the grafted TMSPMA and the acrylate groups in acrylamide under a thermal-induced free radical polymerization. Consequently, long-chain PAAm polymer network was covalently anchored to the glass fixtures, so that we can capture the nonlinear large deformation of hydrogels up to failure of materials rather than failure at the interface between materials and fixtures.

**Measurement of stress-stretch curves and bulk hysteresis.** We used mechanical tester (Zwick/Roell Z2.5) to measure the stress-stretch curves of hydrogels at a fixed loading speed of 10 mm/min. The nominal stress $S$ was measured from the recorded force divided by width and thickness of the sample. The applied stretch $\lambda$ was monitored by the recorded displacement divided by height of the sample. We first performed tensile loading under a monotonic loading, measuring the stress-stretch curves of hydrogels up to the failure of material. We then performed a single cycle of loading and unloading on the other pristine sample with controlled maximum applied stretch $\lambda_{applied}$ from a relatively small value to a high value approaching the ultimate stretch of the material. The bulk hysteresis was calculated by the ratio of the enclosed loop area of loading and unloading stress-stretch curves to the enclosed area of loading curves, namely, $h(\lambda_{applied}) = \oint_1^{\lambda_{applied}} Sd\lambda / \int_1^{\lambda_{applied}} Sd\lambda$. The bulk hysteresis typically increases monotonically with $\lambda_{applied}$, and reaches a maximum plateau value, which was identified as the maximum bulk hysteresis $h_m$.

**Fracture test.** We adopted pure-shear tensile method to measure the fracture energies of the two series of PAAm-alginate hydrogels Fig. S8(a). Given the measured nominal stress versus stretch curves, we can calculate the mechanical work done on the unnotched sample as $U(\lambda_{applied}) = \int_1^{\lambda_{applied}} Sd\lambda$. We further introduced a sharp crack in the other pristine sample. The crack length was controlled about one forth of the width of the sample. We then applied tensile loading on the notched sample at a fixed loading speed of 10 mm/min, measuring a critical stretch $\lambda_c$, at which crack propagates steadily. The measured critical stretches are summarized in Fig. S8(b). Given the measured $\lambda_c$, we can calculate the fracture toughness as $\Gamma = H \int_1^{\lambda_c} Sd\lambda$. The measured fracture energies are summarized in Fig. S8(c).

**Fatigue test.** We adopted both pure-shear and single-notch methods to measure the fatigue thresholds of the two series of PAAm-alginate hydrogels. For the pure-shear method, we fabricated the sample into a rectangular shape with dimensions of 34 × 6 × 1.5 mm³ at its as-prepared state.

The sample was chemically anchored on glass fixtures. As schematically illustrated in Fig. S5a, we first cyclically load an unnotched sample to measure the steady-state nominal stress versus stretch curve under cyclic loading. The strain energy density $W$ of the unnotched sample under the $N^{th}$ cycle of maximum applied stretch of $\lambda^A$ can be calculated as $W(\lambda^A, N) = \int_1^{\lambda^A} S(N) d\lambda$ with $S$ and $\lambda$ being the steady-state nominal stress and stretch, respectively. Thereafter, a cyclic loading with the same maximum stretch of $\lambda^A$ is applied on the other notched sample with the same dimensions as the unnotched sample. The corresponding applied energy release rate can be calculated as $G(\lambda^A, N) = H \int_1^{\lambda^A} S(N) d\lambda$ with $H$ being height of the sample. We used a camera (Imaging Source, 30 μm/pixel) to record the crack extension (i.e., $\Delta C$) over cycles, measuring the crack extension rate $dC/dN$. By systematically varying the applied stretch $\lambda^A$, we can obtain a plot of $dC/dN$ versus $G$. Figure S4(b) shows a representative plot of $dC/dN$ versus $G$ for the hydrogel with $Ca^{2+}$ and alginate concentration of $C_A$ = 1.0 wt%. By linearly extrapolating the curve of $dC/dN$ versus $G$ to the intercept with the abscissa, one can approximately identify the measured fatigue threshold $\Gamma_0$ in the swollen state. Give the swelling ratio in volume $\lambda_V$, one can further calculate the fatigue threshold of the hydrogel in the as-prepared state by $\lambda_V^{2/3} \Gamma_0$ (Fig. S7).

For the single-notch method, we fabricated the sample into a dog-bone shape. As schematically illustrated in Fig. S6a, we first cyclically load an unnotched sample to measure the steady-state nominal stress versus stretch curve under cyclic loading. The strain energy density $W$ of the unnotched sample under the $N^{th}$ cycle of maximum applied stretch of $\lambda^A$ can be calculated as $W(\lambda^A, N) = \int_1^{\lambda^A} S(N) d\lambda$ with $S$ and $\lambda$ being the steady-state nominal stress and stretch, respectively. Thereafter, the same cyclic stretch $\lambda^A$ is applied on the notched sample, measuring the evolution of the cut length in undeformed state $c$ as a function of the cycle number. The initial cut length is smaller than one fifth the width of the sample. The applied energy release rate can be calculated as $G(\lambda^A, N) = 2k(\lambda^A) c(N) \int_1^{\lambda^A} S(N) d\lambda$, where $k = 3/\sqrt{\lambda^A}$ and $c$ is the current crack length at undeformed configuration. We used a camera (Imaging Source, 30 μm/pixel) to record the crack extension (i.e., $\Delta C$) over cycles, measuring the crack extension rate $dC/dN$. By systematically varying the applied stretch $\lambda^A$, we can obtain a plot of $dC/dN$ versus $G$. Figure

S5(c) shows a representative plot of $dC/dN$ versus $G$ for the hydrogel with $Ca^{2+}$ and alginate concentration of $C_A = 1.0$ wt%. By linearly extrapolating the curve of $dC/dN$ versus $G$ to the intercept with the abscissa, one can approximately identify the measured fatigue threshold $\Gamma_0$ in the swollen state. Give the swelling ratio in volume $\lambda_V$, one can further calculate the fatigue threshold of the hydrogel in the as-prepared state by $\lambda_V^{2/3}\Gamma_0$ (Fig. S7).

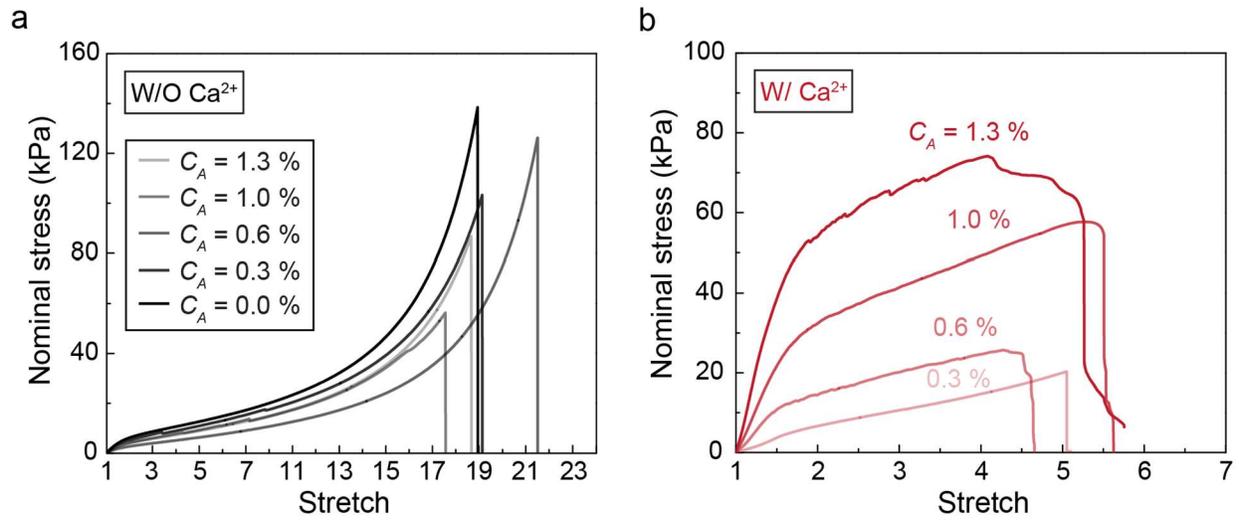

FIG. S1. Nominal stress versus stretch curves of (a) PAAm-alginate hydrogels without $Ca^{2+}$ containing various alginate concentrations $C_A$ and (b) PAAm-alginate hydrogels with $Ca^{2+}$ containing various alginate concentrations $C_A$.

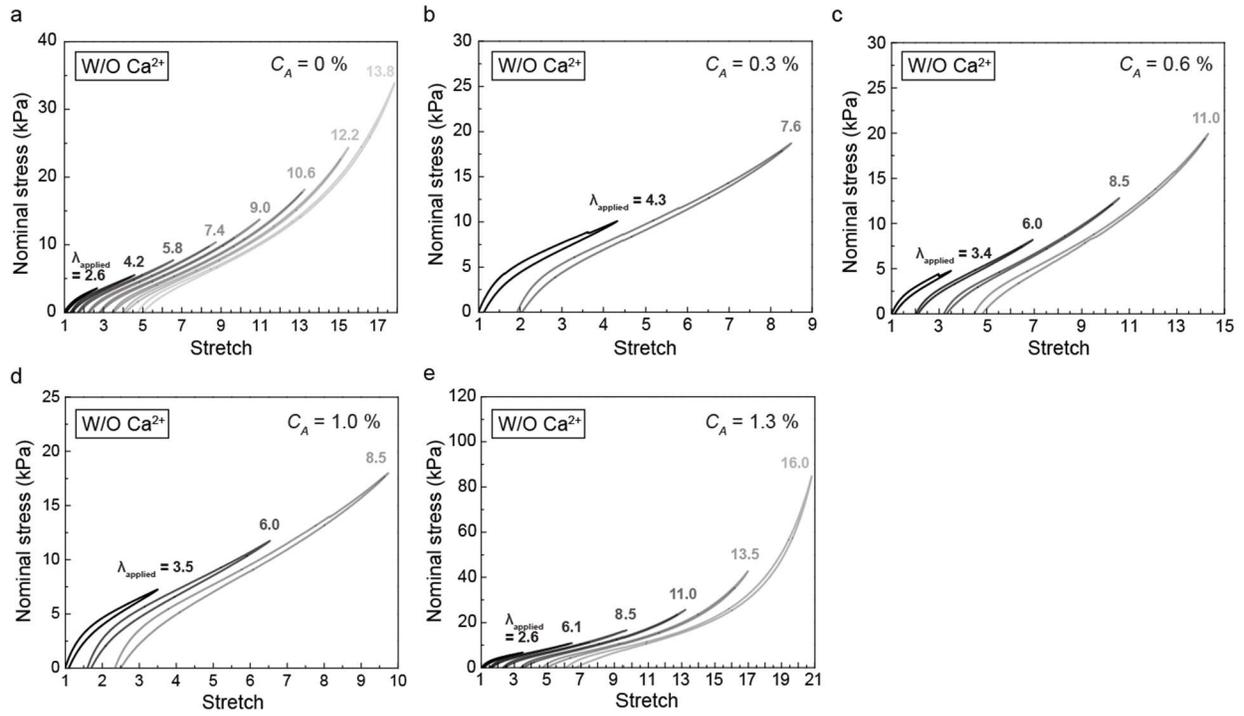

FIG. S2. Nominal stress versus stretch curves of PAAm-alginate hydrogels without $Ca^{2+}$ under one cycle of loading and unloading. (a) $C_A$ = 0.0 wt%. (b) $C_A$ = 0.3 wt%. (c) $C_A$ = 0.6 wt%. (d) $C_A$ = 1.0 wt%. (e) $C_A$ = 0.3 wt%.

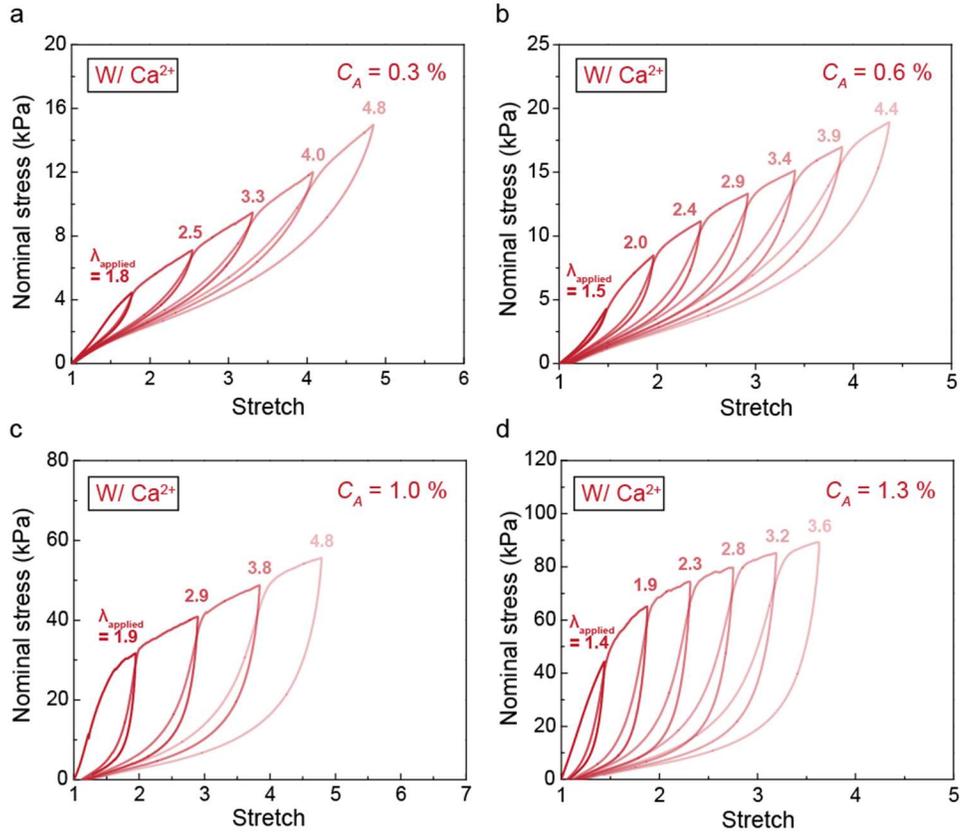

FIG. S3. Nominal stress versus stretch curves of PAAm-alginate hydrogels with $Ca^{2+}$ under one cycle of loading and unloading. (a) $C_A$ = 0.3 wt%. (b) $C_A$ = 0.6 wt%. (c) $C_A$ = 1.0 wt%. (d) $C_A$ = 1.3 wt%.

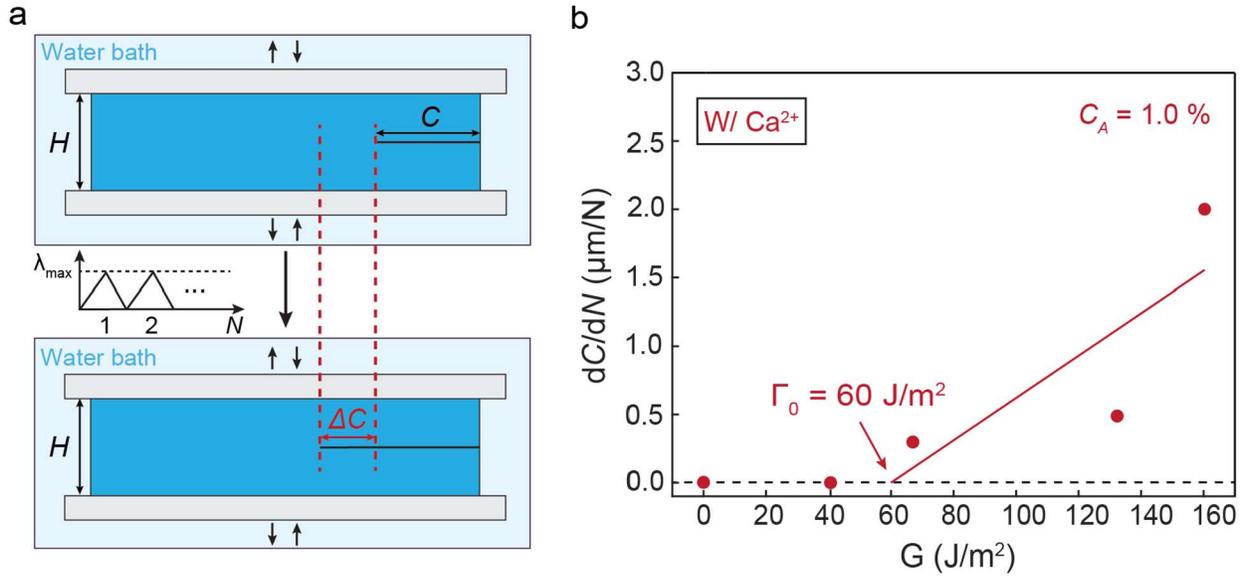

FIG. S4. Measurement of fatigue-threshold using pure-shear tensile tests. (a) Schematic illustration of the fatigue characterization using pure-shear tensile tests. (b) A representative crack extension curve in the plot of crack extension rate (i.e., d$C$/d$N$) as a function of applied energy release rate G (J/m$^2$). The fatigue threshold is identified as $\Gamma_0 = 60$ J/m$^2$ for the hydrogel with Ca$^{2+}$ and alginate concentration of $C_A = 1.0$ wt%.

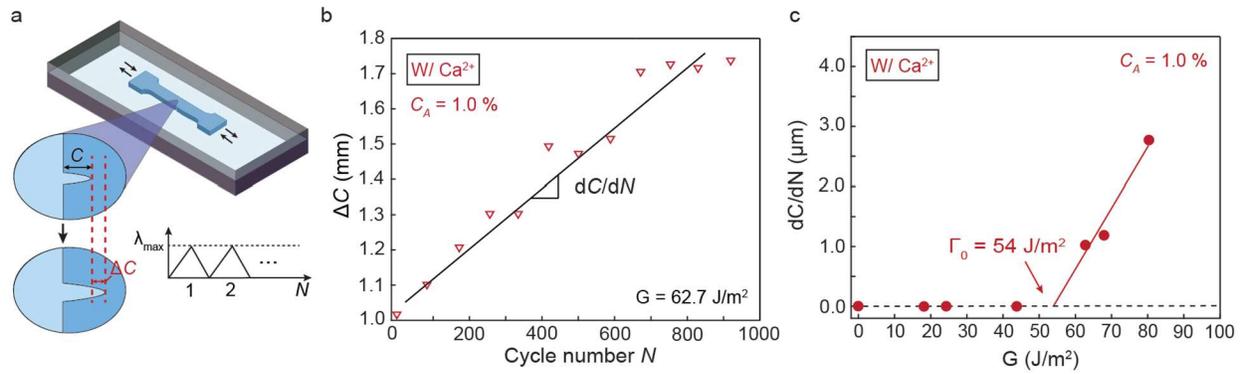

FIG. S5. Measurement of fatigue-threshold using single-notch tensile tests. (a) Schematic illustration of the fatigue characterization using single tensile tests. (b) A representative crack extension $\Delta C$ versus cycle number $N$ for the hydrogel with $Ca^{2+}$ and alginate concentration of $C_A$ = 1.0 wt% at the applied energy release rate of $G = 62.7$ J/m$^2$. (c) A representative crack extension curve in the plot of crack extension rate (i.e., $dC/dN$) as a function of applied energy release rate $G$ (J/m$^2$). The fatigue threshold is identified as $\Gamma_0 = 54$ J/m$^2$ for the hydrogel with $Ca^{2+}$ and alginate concentration of $C_A$ = 1.0 wt%.

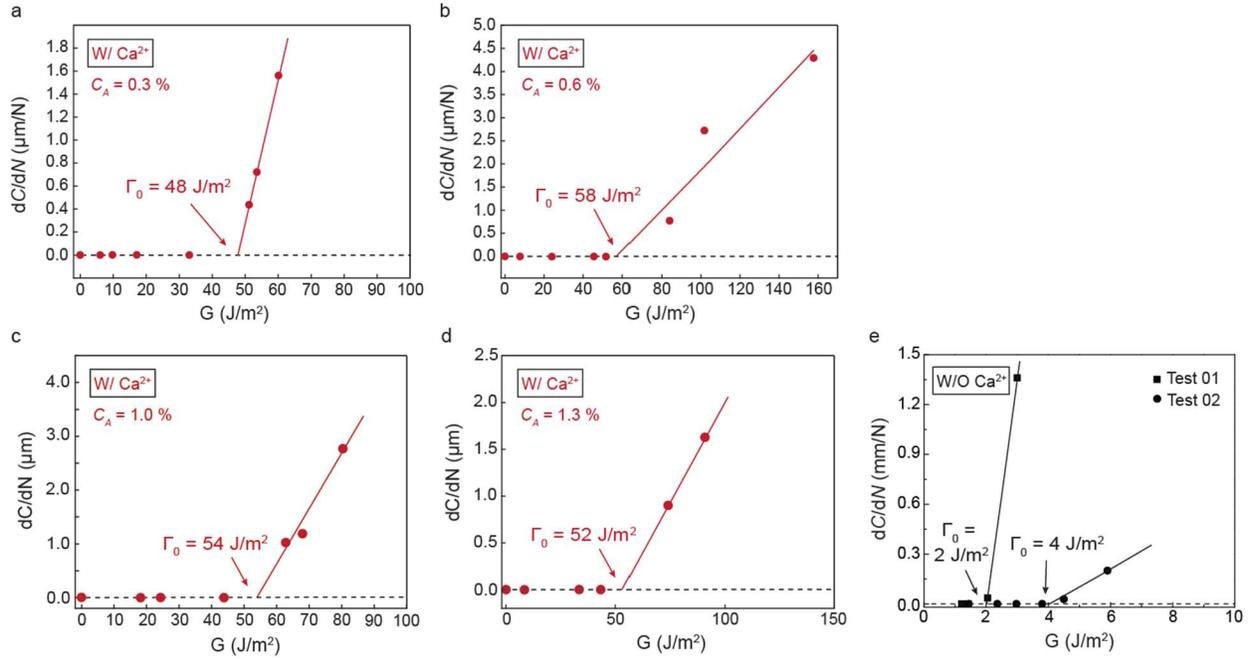

FIG. S6. Summarized fatigue-induced crack extension curves. (a) Hydrogels with $Ca^{2+}$ and alginate concentration $C_A = 0.3$ wt%. (b) Hydrogels with $Ca^{2+}$ and alginate concentration $C_A = 0.6$ wt%. (c) Hydrogels with $Ca^{2+}$ and alginate concentration $C_A = 1.0$ wt%. (d) Hydrogels with $Ca^{2+}$ and alginate concentration $C_A = 1.3$ wt%. (e) Hydrogels with no alginate or no $Ca^{2+}$.

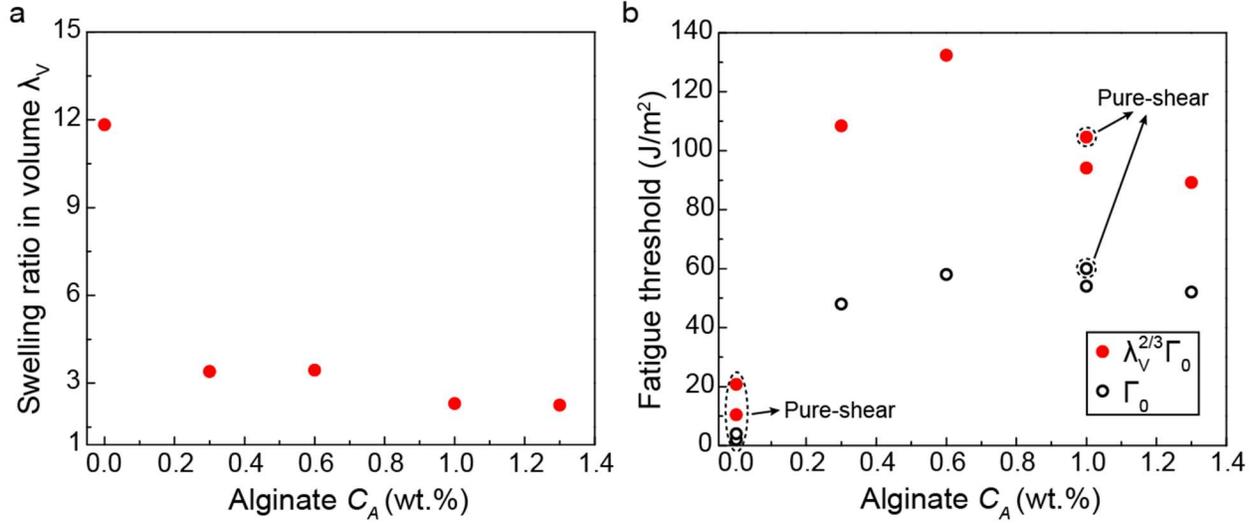

FIG. S7. (a) Swelling ratio in volume of the hydrogels $\lambda_V$ with various alginate concentrations soaking in a bath of 0.01 wt% calcium chloride. (b) Summarized fatigue thresholds of hydrogels with various alginate concentrations in the swollen state $\Gamma_0$ (hollow dots) and in the as-prepared state $\lambda_V^{2/3}\Gamma_0$ (solid dots). The circled data are measured using pure-shear method and the other data are measured using single-notch method.

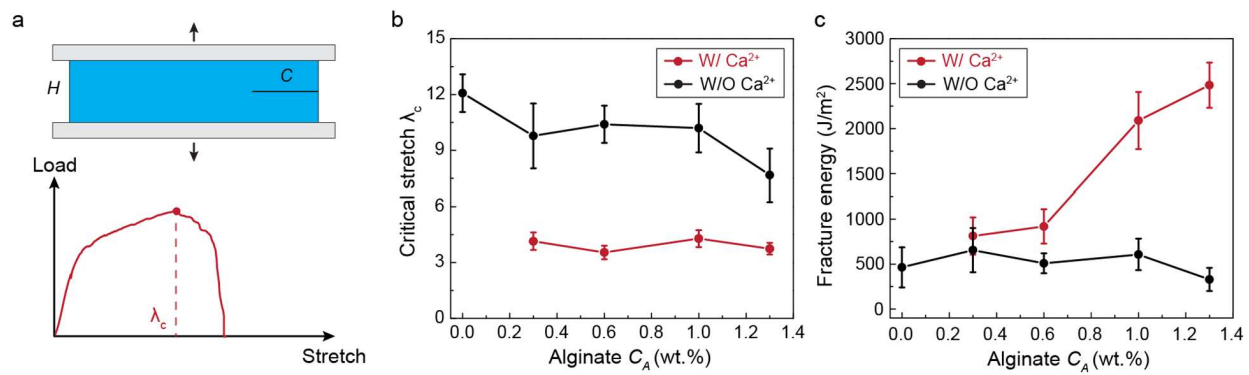

FIG. S8. Fracture test to measure the fracture toughness of the two series of PAAm-alginate hydrogels. (a) Schematic illustration of the pure shear tensile test for fracture test. (b) Summarized critical stretches. (c) Summarized fracture energies.